\documentclass[twocolumn]{article}
\usepackage[left=0.7in, right=0.7in, top=1.1in, bottom=1.1in]{geometry}
\usepackage{graphicx} 
\usepackage{hyperref}
\usepackage{soul, xcolor}
\usepackage{amsfonts, amsmath, amssymb}
\hypersetup{
    colorlinks=true,
    linkcolor=blue,    
    citecolor=blue,    
    urlcolor=blue,     
}

\title{SAR and InSAR Change Detection with Quantum Generative Models}

\usepackage{authblk}

\author[1]{\mbox{Samwel K. Sekwao}}
\author[2]{\mbox{Shaunak De}}
\author[1]{\mbox{Alexis Hocken}}
\author[2]{\mbox{Scott Staniewicz}}
\author[1]{\mbox{Evgeny Epifanovsky}}
\author[2]{\mbox{Craig Stringham}}
\author[2]{\mbox{Gordon Farquharson}}
\author[1]{\mbox{Martin Roetteler}}
\author[1]{\mbox{Panagiotis Kl. Barkoutsos}}
\author[1]{\mbox{Jason Iaconis}}
\affil[1]{IonQ Inc., 4505 Campus Drive, College Park, MD 20740, USA}
\affil[2]{Capella Space (an IonQ company), 438 Shotwell Street, San Francisco, CA 94110, USA}

\date{\today}

\begin{document}

\maketitle

\begin{abstract}

Change detection in synthetic aperture radar (SAR) and interferometric synthetic aperture radar (InSAR) underpins disaster response, infrastructure monitoring and land-use enforcement. 
Detection is limited by the background estimator, which conventionally forms a conditional expectation directly from observed pixel statistics and degrades where those statistics are sparse, including the regime produced by the heavy-tailed marginals of sub-meter-resolution radars. 
In this work, we integrate state-of-the-art satellite imagery with quantum machine learning on IonQ trapped-ion-based quantum processors. By replacing the empirical conditional with a quantum circuit Born machine (QCBM)-sampled generative model in Copula space, we substantially improve change detection on sparse real-world images.
On Capella Space satellite image acquisitions, the generative estimator matches conventional methods when the observed statistics are adequate, and substantially outperforms them when they are not. 
Executing the trained model on IonQ trapped-ion based hardware reproduces the results of the ideal and noisy simulations and demonstrates up to par, or even better, performance with the classical state-of-the-art methods. 
For a SAR dataset of an airport, QPU circuit evaluations for both training and inference achieved a maximized filtered F1 score of 0.32, compared with 0.16 and 0.24 for the two classical baselines. For an InSAR dataset of a volcanic lava flow, all three methods reached a maximum filtered F1 of approximately 0.66. These experiments demonstrate the feasibility of executing a QCBM-based background estimator on trapped-ion hardware. We further demonstrate that the QCBM method successfully extends to interferometric coherence data, achieving performance comparable to classical approaches.
\end{abstract}

\section*{Main Text}
Synthetic aperture radar (SAR) change detection identifies regions of a scene that have altered between two or more acquisitions of the same area, typically collected at different times by the same or comparable sensors. Because SAR is an active, coherent imaging modality, it operates independently of solar illumination and largely independently of cloud cover, making it a persistent, all-weather, day-and-night complement to optical change detection for monitoring dynamic environments \cite{henderson1997sar}. Two broad families of methods dominate operational SAR change detection, distinguished by which component of the complex-valued SAR signal they exploit. Amplitude (or intensity) change detection compares the backscattered power recorded in each acquisition, typically through a ratio or log-ratio operator on coregistered image pairs, followed by automatic or statistical thresholding of the resulting difference image to separate changed from unchanged pixels \cite{touzi1988statistical, bruzzone2000automatic, bazi2005unsupervised, bovolo2005detail}. More recent work replaces the fixed operator and threshold with learned representations, classifying change directly from the bi-temporal pair using deep networks \mbox{\cite{gong2017feature}}.
\par
Amplitude change detection relies solely on the back scattered power of the radar return, without requiring the phase information recorded by the sensor. Because it does not depend on maintaining phase coherence between passes, this approach can be applied across a wider range of acquisition geometries and does not require the tight orbital control needed to form a stable interferometric baseline. Its principal limitation is that backscatter intensity is strongly dependent on the aspect angle at which a target is illuminated, so changes in backscatter driven by variation in viewing geometry, rather than by a genuine change in the scene, can be misclassified as detected change.
\par
Interferometric SAR (InSAR) change detection instead exploits the phase of two complex SAR acquisitions, either through the interferometric phase difference itself to measure millimeter-to-centimeter scale ground movement, or through the loss of coherence between the two acquisitions, which reveals disturbance of the scattering surface at scales well below the resolution of the sensor \cite{Bamler1998SyntheticApertureRadar,quach2015model}. It is therefore sensitive to changes that leave the backscattered amplitude essentially unaltered, such as subsidence, structural deformation and vehicle tracks or other surface disruption. This sensitivity comes at the cost of a far more demanding acquisition requirement. The two images must be collected from nearly the same orbital position and incidence angle, and co-registered to a small fraction of a resolution cell \cite{plyer2015new}.
\par
In practice, amplitude and interferometric change detection are therefore complementary rather than competing techniques, with the choice, or combination, of the two governed by the type of change to be detected and the acquisition geometry available. These characteristics make SAR change detection central to national-security and civil applications alike. For defense and intelligence applications, the ability to reliably detect man-made and natural changes over wide areas, at high resolution, and independent of weather or lighting conditions, directly supports activity monitoring, treaty verification and wide-area search for objects or infrastructure of interest \cite{quach2015model, NLCD0}. The same capability underlies a broad range of civil applications, including disaster response and damage assessment, monitoring of the built environment and settlement growth \cite{henderson1997sar}, and land-use enforcement, all of which depend on detecting genuine scene changes against a background of sensor noise, seasonal variation and imperfect coregistration. Accordingly, we evaluate the method developed here on one dataset of each type: a high-resolution amplitude pair, and a pair of interferometric coherence maps.

\par
Across both families, the detection problem reduces to the same statistical task: estimating what the second acquisition \emph{would} have looked like had the scene not changed, and flagging the pixels that depart from it. Under a mean-squared-error criterion the optimal such background estimator is the conditional expectation of the observed intensity given the reference intensity, $E[y \mid x]$ \cite{NLCD0}, and in practice this conditional is formed empirically, by accumulating a joint histogram of coregistered pixel pairs and reading off the mean of each conditioning bin. The quality of the resulting background image is therefore governed not by the sophistication of the estimator but by how many observations populate each bin. However, this ceases to be a good approach at high resolution. Single-look amplitude statistics of sub-meter X-band imagery are strongly skewed and heavy-tailed, and are poorly described by the homogeneous-clutter models that adequately represent coarser-resolution scenes \cite{frery1997model}, so the occupied region of the joint histogram is a narrow, sparsely sampled ridge rather than a well-filled cloud with an average occupancy well below one sample per bin. Conditioning bins are consequently empty or populated by a handful of samples, and the conditional means read from them are correspondingly noisy or undefined. The degradation is worst in the distribution tails, which is exactly where detection at low false-alarm rates operates, and it compounds the nuisance variability already present from aspect-angle dependence, seasonal change and residual coregistration error. The conventional remedy is to regularize the statistics before estimation, by applying variance-stabilizing or Gaussianizing transformations and spatial filtering until the histogram is dense enough to support a lookup table, at the cost of the fine spatial detail that motivated acquiring high-resolution imagery in the first place. The alternative pursued here is to leave the data alone and replace the empirical conditional with one drawn from a generative model of the joint distribution, which can be sampled without limit and so supplies well-defined conditionals in regions the observations never populated. We model that joint distribution through its copula, which separates the dependence structure between the two acquisitions from their individual marginals, and which has an established precedent in remote-sensing change detection \cite{mercier2008conditional}. However, training generative models on images with sparsely populated bins poses significant practical challenges.

\par
Quantum computing exploits superposition, entanglement, and interference to process information in ways that differ fundamentally from classical computation, offering the potential for computational advantages on specific classes of problems\cite{harrow2017quantum,daley2022practical}. Recent advances in quantum hardware, control, and quantum error correction are progressively improving the scale and fidelity of quantum processors, while hybrid quantum–classical approaches provide a practical framework for exploring applications on current devices\cite{debry2026error,barthe2025parameterized}. Quantum machine learning has consequently emerged as an active area of research, with parameterized quantum circuits being investigated for classification, regression, and generative modeling, including recent demonstrations of their ability to represent multivariate probability distributions\cite{barthe2025parameterized}. Among these approaches, quantum circuit Born machines (QCBMs) are generative models in which a parameterized quantum circuit encodes a probability distribution and measurements of the trained circuit generate samples from that distribution\cite{QCBM0,QCBM1,QCBM2}. Their ability to learn and sample correlated multivariate distributions makes QCBMs particularly relevant to image change detection, where the statistical dependence between corresponding observations in before and after images provides information for estimating the unchanged background.

\par
In this work we replace the empirical conditional expectation at the heart of the background estimator with one sampled from a QCBM trained on the joint distribution of before- and after-image intensities in copula space, and we test the resulting detector against the classical estimators it displaces. We evaluate on two bi-temporal Capella Space acquisitions: an X-band Stripmap amplitude pair at $1.2\,\mathrm{m}$ resolution over Marine Corps Air Station Miramar, San Diego, and a pair of interferometric coherence maps spanning the 2026 eruptive sequence of Piton de la Fournaise. As baselines we use the nonlinear background estimator (NLBE), which forms $E[y \mid x]$ from a lookup table over the observed joint histogram, and a classical copula variant of the same estimator. Across the evaluated datasets, we find that QCBM outperformed the classical baselines for the more challenging image pairs characterized by non-Gaussian intensity distributions. In contrast, for datasets with approximately Gaussian intensity distributions, results show that QCBM and classical estimators achieved comparable change-detection performance.

\paragraph*{A generative estimator for conditional background modeling.} 

 When QCBMs are constructed using quantum circuits with specific structural constraints, their outputs are entangled quantum states which can naturally represent copulas. This provides a natural framework for capturing statistical dependencies between variables. Recent studies have applied such QCBMs to financial data with promising results, including learning joint distributions of historical stock-market data and modeling multivariate copulas for financial risk aggregation \cite{PhysRevResearch.4.043092,Zhu2023Copula}. Motivated by these applications, we formulate image change detection as a generative modeling problem in which a QCBM learns the joint distribution of corresponding pixel intensities in the before and after images and uses the learned distribution to estimate the expected background.

\begin{figure*}[h!]
    \centering
    \includegraphics[width=\textwidth]{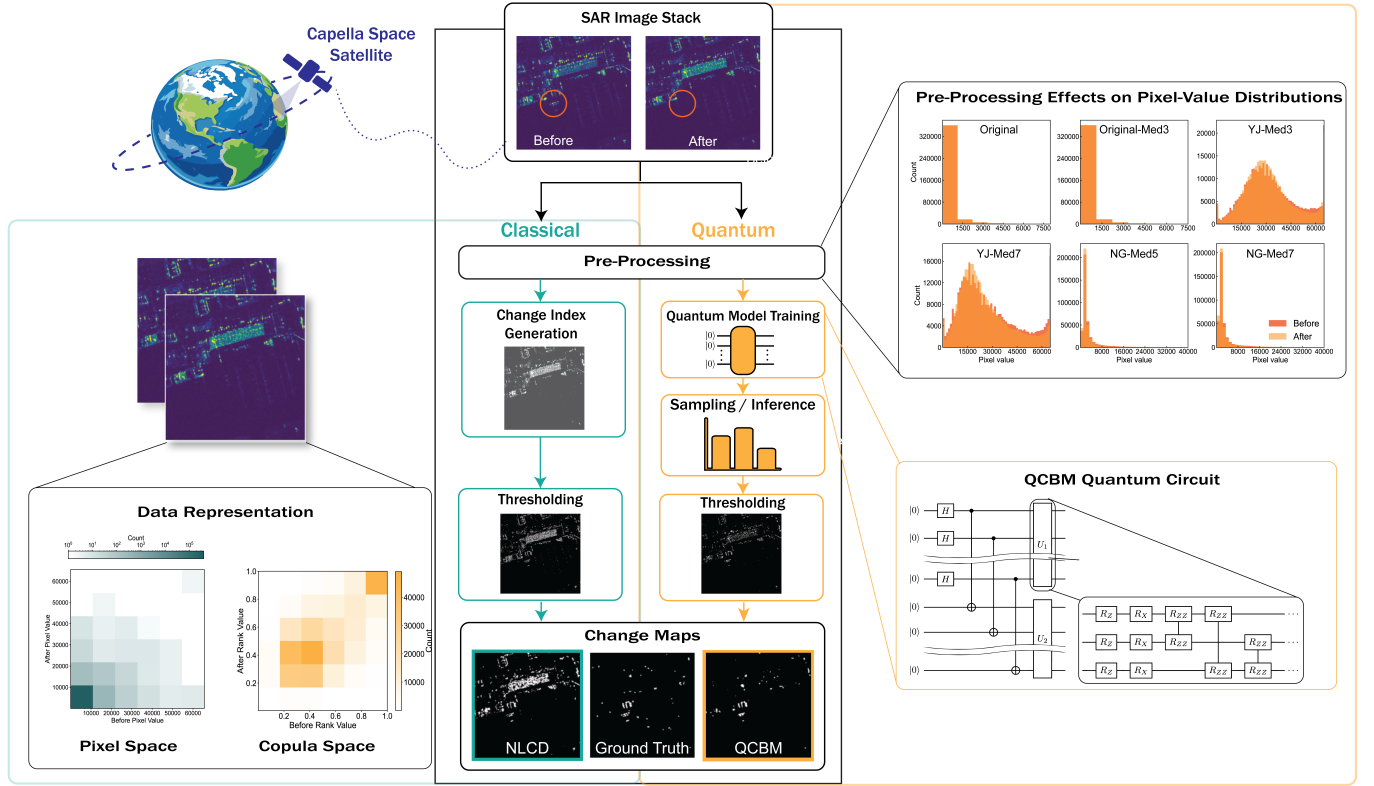}
    \caption{Overview schematic of classical and quantum pipelines for processing a given before and after SAR image stack acquired from Capella Space Satellite. Pixel-value distributions of the before and after images for the different preprocessing configurations of the Miramar1 dataset. The original Miramar1 and median-filtered (Med)/non-Gaussian (NG)transformed datasets exhibit strongly non-Gaussian distributions, whereas the Yeo–Johnson (YJ) transformed datasets exhibit more Gaussian-like distributions. The histograms and heatmaps illustrate the effect of preprocessing on the statistical characteristics of the before and after image pixels. The quantum circuits used in the QCBM method are also illustrated.}
    \label{fig:miramar1_all_images}
\end{figure*}

For image change detection, the QCBM uses two $n_{bits}$-qubit registers to represent the discretized before- and after-image intensities, requiring $2n_{bits}$ qubits in total, as illustrated by the circuit architecture in Figure \ref{fig:miramar1_all_images}. This circuit first prepares a state with maximal entanglement between the two registers, ensuring that each marginal distribution of each variable is uniformly distributed, as in copulas. Parameterized quantum gates are then applied, which enable the circuit to learn dependencies between the two image variables.  The pixel intensities are transformed into copula space using their marginal CDFs and discretized to define the target distribution. The QCBM is then trained by minimizing the Kullback–Leibler (KL) divergence between the target and circuit-generated distributions using a classical optimizer.

Following training, measurements of the QCBM generate samples that are transformed from copula space back to the original pixel-intensity domain. These samples are conditionally selected based on the observed image intensity, and their conditional mean is used to estimate the corresponding background intensity. Repeating this process generates the background image for change detection. Unlike classical approaches that estimate conditional distributions directly from the observed data, the QCBM obtains them from samples generated by the learned quantum model.

\paragraph*{Change detection performance on high-resolution SAR and InSAR images.}

The Miramar dataset comprises paired before and after images acquired at three distinct locations within the Miramar airport in San Diego, CA. The image pairs corresponding to the first, second, and third locations are denoted as datasets Miramar1, Miramar2, and Miramar3, respectively. The quantum image change detection method was evaluated on the Miramar1 dataset and compared with the classical methods NLBE and Copula described in the Methods section. Following the naming convention adopted in this work, NLBE is hereafter referred to as Non-Linear Change Detection (NLCD). To examine model performance under different image-intensity distributions, Miramar1 was evaluated using several preprocessing configurations designed to produce both non-Gaussian and approximately Gaussian pixel-value distributions, as illustrated in Figure \ref{fig:miramar1_all_images}. The unprocessed Miramar1 image pair is referred to as Original, while Original-Med3 applies a $3\times3$ median filter to the original images. YJ-Med3 and YJ-Med7 apply a Yeo–Johnson (YJ) transformation followed by $3\times3$ and $7\times7$ median filtering, respectively. In contrast, NG-Med5 and NG-Med7 apply a non-Gaussian (NG) transformation followed by $5\times5$ and $7\times7$ median filtering, respectively. Together, these configurations enable evaluation of the quantum and classical change detection methods across different preprocessing conditions and pixel-value distribution characteristics.

The Original and Original-Med3 datasets retained strongly skewed, long-tailed pixel distributions, whereas YJ-Med3 and YJ-Med7 produced more Gaussian-like distributions after the Yeo-Johnson transformation. The NG-Med5 and NG-Med7 configurations provided additional non-Gaussian cases. Across these configurations, the before- and after-image marginal distributions showed substantial overlap, indicating that performance depends on modeling the joint distribution of before- and after-image pixel values rather than on marginal separation alone.

\begin{table}[h!]
\caption{Comparison of the best filtered F1 scores obtained by QCBM, NLCD, and Copula on Gaussian and non-Gaussian preprocessing configurations of the Miramar1 dataset. QCBM training and inference were performed using an ideal quantum simulator. Bold values denote the best-performing method for each non-Gaussian dataset.\\}
\centering
\begin{tabular}{ c c c c c } 
 \bf{Dataset} &  \bf{QCBM} & \bf{NLCD} & \bf{Copula}\\ 
 \hline
 Original & \bf{0.30} & 0.14 & 0.20 \\ 
 Original-Med3 & \bf{0.41} & 0.16 & 0.24 \\ 
 YJ-Med3 & 0.56 & 0.55 & 0.56\\ 
 YJ-Med7 & 0.54 & 0.53 & 0.54\\ 
 NG-Med5 & \bf{0.36} & 0.21 & 0.26\\ 
 NG-Med7 & \bf{0.32} & 0.25 &  0.29\\ 
 \hline
\end{tabular}
\label{table:1}
\end{table}

Table \ref{table:1} reports the best F1 scores obtained after applying the uniform filtering algorithm to the change masks generated by QCBM, NLCD and Copula. Uniform filtering was performed using the  \texttt{uniform\_filter} function from \texttt{scipy.ndimage}\cite{Scipy}, which replaces each pixel with the local mean within a specified neighborhood. When applied to a binary change mask, the filtered value therefore represents the local fraction of pixels classified as changed. The filtered output was subsequently thresholded to retain spatially coherent change regions and suppress isolated detections. Further details of the uniform-filtering procedure and parameter selection are provided in the Methods section. For the results in Table \ref{table:1}, both QCBM training and inference were performed using an ideal quantum simulator. On the non-Gaussian datasets, QCBM consistently achieved the highest F1 score after uniform filtering. For the Original Miramar1 dataset, QCBM reached 0.30, compared with 0.14 for NLCD and 0.20 for Copula. After 3 x 3 median filtering, QCBM improved to 0.41 on Original-Med3, whereas NLCD and Copula reached 0.16 and 0.24, respectively. The same trend was observed for NG-Med5 and NG-Med7, where QCBM achieved 0.36 and 0.32, compared with 0.21 and 0.25 for NLCD and 0.26 and 0.29 for Copula. In contrast, when the Yeo-Johnson transformation produced approximately Gaussian-distributed data, the performance gap largely disappeared: on YJ-Med3, QCBM and Copula both reached 0.56 and NLCD reached 0.55, while on YJ-Med7, QCBM and Copula both reached 0.54 and NLCD reached 0.53. These results show that the classical methods benefit strongly from preprocessing that regularizes the pixel distributions, whereas QCBM retains a relative performance advantage on the more challenging non-Gaussian configurations.

\begin{figure}[h!]
    \centering
    \includegraphics[width=\columnwidth]{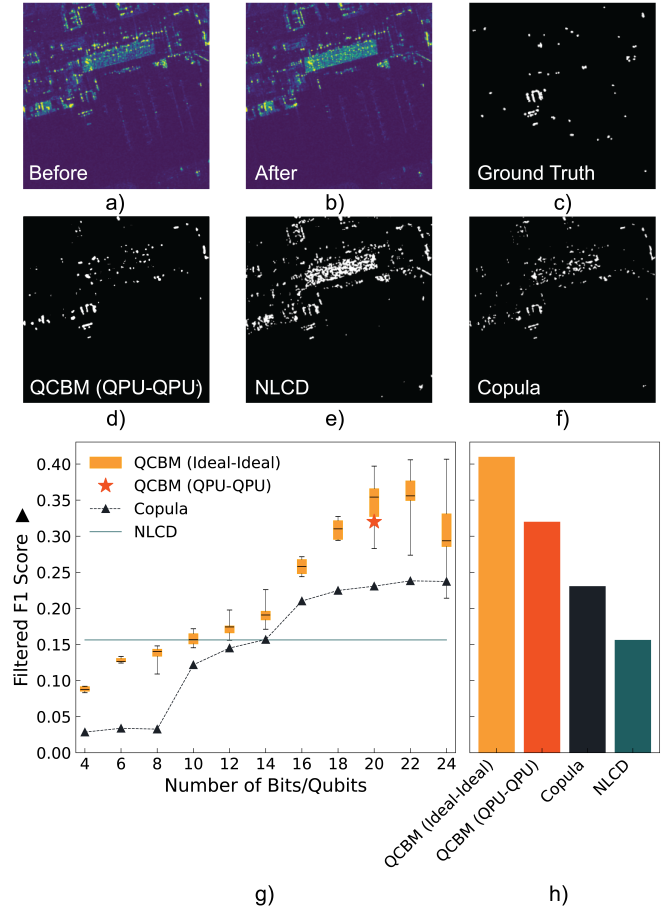}
    \caption{Qualitative comparison of change-detection results for the non-Gaussian Miramar1-Med3 dataset. The top row shows the a) before image, b) after image, and c) ground-truth change mask, while the middle row shows the masks generated by d) QCBM, e) NLCD, and f) Copula. The QCBM training and inference were both performed on IonQ's Forte-Enterprise QPU. g) Filtered F1 scores for QCBM (training and inference performed with an ideal simulator), Copula, and NLCD at varying number of bits vs qubits for the Miramar1-Med3 dataset. h) Comparison of the best F1 scores for models trained with 20 bits/qubits.}
    \label{fig:miramar1_med3_masks}
\end{figure}

The qualitative comparison for the non-Gaussian Miramar1-Med3 dataset is shown in Figure \ref{fig:miramar1_med3_masks}a-f. The selected QCBM used 10 bits per variable, corresponding to 20 qubits in total, with 10 bits assigned to the before-image pixel values and 10 bits assigned to the after-image pixel values. The QCBM model's change mask in Figure \ref{fig:miramar1_med3_masks}d represents the output when both QCBM training and inference were performed on IonQ's Forte-Enterprise QPU. The best-performing Copula configuration used 11 bits per variable. Under QPU inference, QCBM achieved a filtered F1 score of 0.37, compared to 0.16 for NLCD and 0.24 for Copula.

The QCBM mask was sparser than the masks generated by the classical methods and more closely followed the spatial distribution of changes in the ground truth. In particular, QCBM reduced background detections in unchanged regions while retaining prominent changed structures. NLCD produced a denser mask with substantial detections outside the ground-truth change regions, and Copula reduced some of this response but still retained more background detections than QCBM. These qualitative differences are consistent with the filtered F1 scores and indicate that the performance trend observed under ideal-simulator evaluation is retained when QCBM inference is transferred to quantum hardware, although the QPU-inference score for Original-Med3 is lower than the corresponding ideal-simulator score reported in Table \ref{table:1}.

\begin{figure}[h!]
    \centering
    \includegraphics[width=\columnwidth]{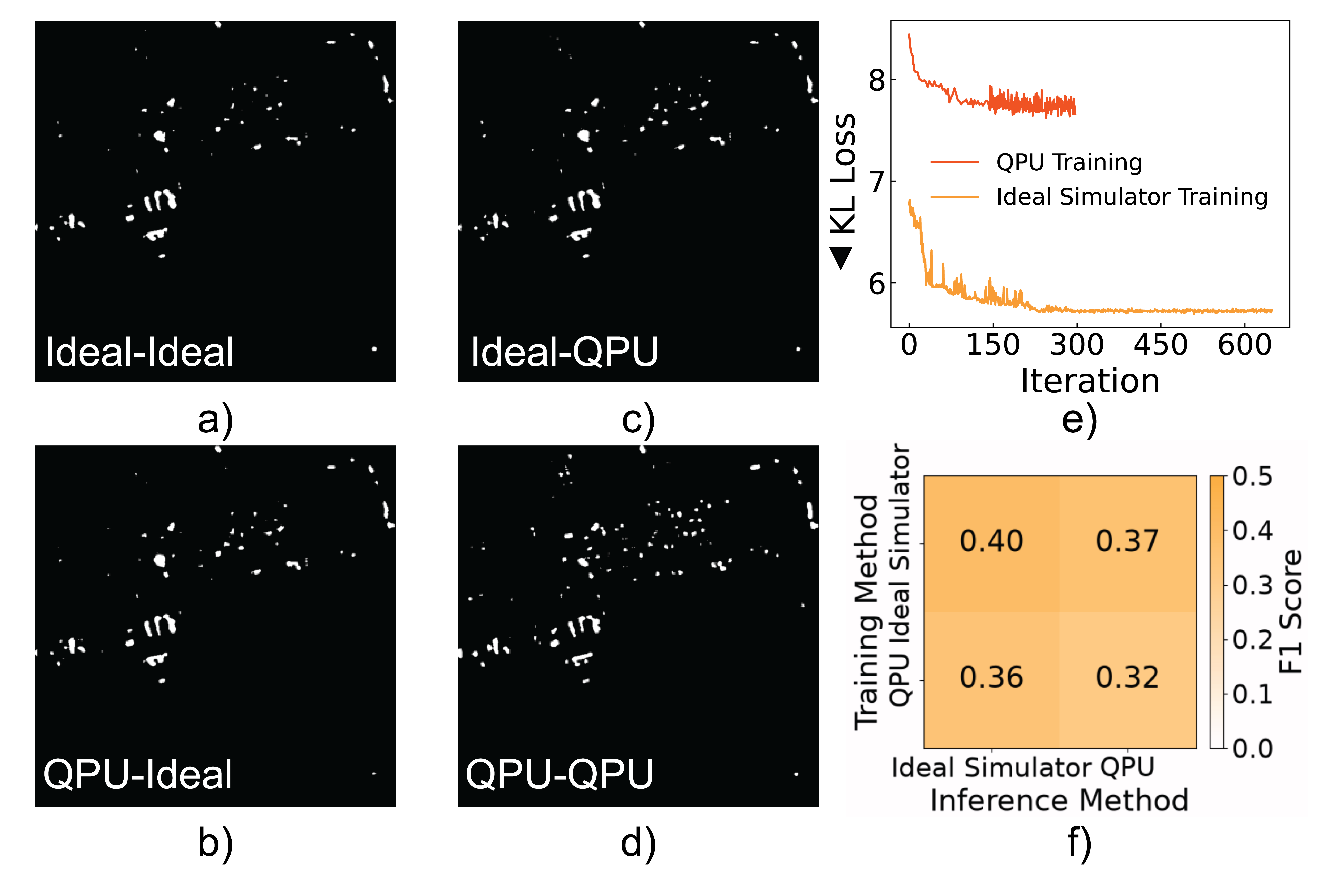}
    \caption{Effect of QPU execution on QCBM training and inference for image change detection on Miramar1-Med3 dataset. The change masks and corresponding filtered F1 scores compare four combinations of ideal-simulator and QPU training and inference (a-d). For QPU execution, training was performed on IonQ's Forte QPU, whereas inference was performed on IonQ's Forte-Enterprise QPU. e) The KL-loss curves compare the convergence behavior of ideal-simulator and QPU training, while f) the heatmap summarizes the filtered F1 scores across the four configurations.}
    \label{fig:miramar_median3_qpu_inference5}
\end{figure}

To further investigate the impact of quantum hardware noise on QCBM-based image change detection, we evaluated the model on Miramar1-Med3 dataset under different combinations of ideal-simulator and QPU execution during training and inference. The ideal-simulator training and inference result was previously reported in Table \ref{table:1}, while the fully hardware-executed QPU training and inference result was discussed in Figure \ref{fig:miramar1_med3_masks}. The experiment in Figure \ref{fig:miramar_median3_qpu_inference5} extends these results by considering the intermediate configurations in which only training or inference is performed on the QPU, thereby allowing the effects of hardware execution at the two stages to be examined separately. The ideal-simulator configuration provides the reference performance, with a filtered F1 score of 0.40. When the ideally trained model was instead evaluated on the QPU, the filtered F1 score decreased only modestly to 0.37, indicating relatively limited sensitivity to QPU noise during inference. In contrast, training on the QPU and subsequently performing inference using the ideal simulator resulted in a larger decrease to 0.32. The fully QPU-based result from Figure \ref{fig:miramar1_med3_masks}, included in Figure \ref{fig:miramar_median3_qpu_inference5} for comparison, achieved a filtered F1 score of 0.32. Thus, changing inference from the ideal simulator to the QPU produces an absolute reduction of approximately 0.03 for either training condition, whereas replacing ideal training with QPU training produces a larger reduction of approximately 0.08 for either inference condition. The corresponding masks in the figure show a consistent qualitative trend, with QPU-trained models exhibiting more apparent false-positive detections. Despite these hardware-related reductions, the QCBM continues to outperform the classical baselines on the dataset. As reported in Table \ref{table:1}, NLCD and Copula achieve filtered F1 scores of 0.16 and 0.24, respectively. In comparison, the two intermediate hardware configurations achieve 0.37 for ideal training/QPU inference and 0.36 for QPU training/ideal inference, while even the fully QPU-executed model discussed earlier achieves 0.32. These results demonstrate that the QCBM's performance advantage over the classical methods is maintained when QPU execution is introduced during either or both stages of the modeling pipeline.

The training-loss curves in Figure \ref{fig:miramar_median3_qpu_inference5}e provide additional context for the larger performance degradation associated with QPU training. Ideal-simulator training, corresponding to the result reported in Table \ref{table:1}, was performed using 100,000 shots per iteration and exhibits a lower and more stable KL loss, whereas QPU training used 5,000 shots per iteration and exhibits greater fluctuations and a higher final loss. The difference in training behavior should not be attributed exclusively to hardware noise because the two training configurations also differ substantially in their number of measurement shots. A lower shot count increases statistical uncertainty in the sampled QCBM probability distribution and consequently introduces additional variability into the KL-divergence objective supplied to the classical optimizer. The degradation associated with QPU training may therefore reflect the combined effects of quantum hardware noise and finite-shot sampling noise. In contrast, the relatively small decrease associated with transferring an ideally trained model to QPU inference suggests that the inference stage is comparatively robust to hardware execution. Further experiments using matched shot counts for ideal and QPU training would be required to isolate the contribution of hardware noise from finite-shot effects and determine their respective impacts on QCBM training.

Figure~\ref{fig:miramar1_med3_masks}(g) examines how discretization resolution affects performance on the non-Gaussian Miramar1-Med3 dataset. For QCBM, the filtered F1 scores represent the mean over 10 independent training runs, and the error bars indicate the standard error of the mean. The same number of bits was used to discretize the before- and after-image pixel values; therefore, a QCBM with $n_{bits}$ bits per variable requires $2n_{bits}$ qubits. The models evaluated in Figure~\ref{fig:miramar1_med3_masks}(g) ranged from 4 qubits at 2 bits per variable to 24 qubits at 12 bits per variable.

The mean QCBM filtered F1 score increased with discretization resolution, rising from approximately 0.09 at 2 bits per variable to a maximum of approximately 0.36 at 11 bits per variable, corresponding to 22 qubits, before decreasing modestly at 12 bits per variable. QCBM began to outperform the bit-independent NLCD baseline, which achieved approximately 0.16, at around 6 bits per variable, corresponding to 12 qubits. Copula also improved with increasing discretization resolution, rising from approximately 0.03 at low bit counts to approximately 0.24 at 11 bits per variable and surpassing NLCD at around 8 bits per variable. However, QCBM outperformed Copula across the full range of discretization resolutions. At 11 bits per variable, QCBM achieved approximately 0.36, compared with 0.24 for Copula and 0.16 for NLCD. Thus, increasing resolution improves both discretized models, but the larger QCBM circuits produced the strongest performance gains on this non-Gaussian SAR dataset.

\begin{figure}[h!]
    \centering
    \includegraphics[width=\columnwidth]{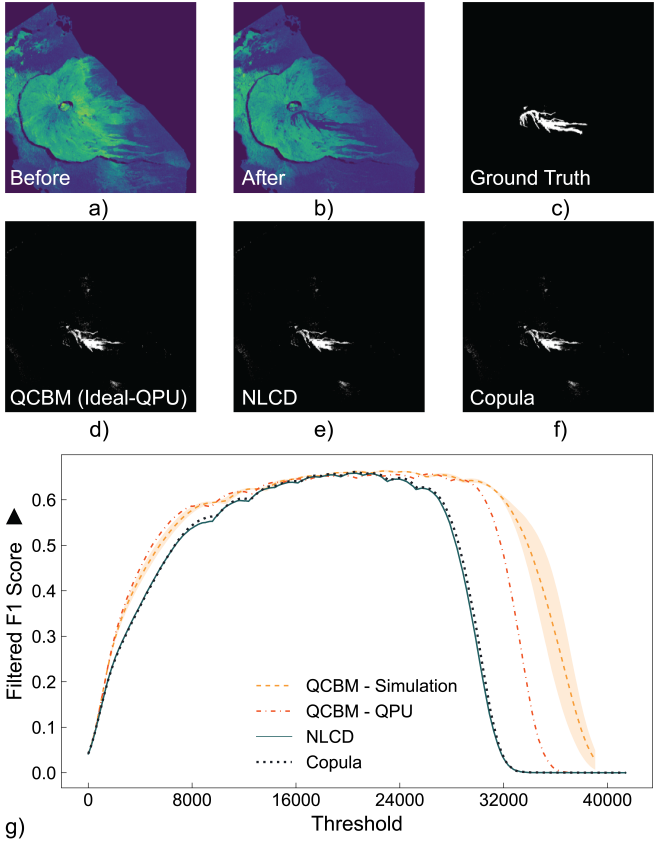}
    \caption{InSAR volcano lava flow change detection with QCBM and classical baselines. a) Before and b) after InSAR images are shown with the c) ground-truth change mask and the masks generated by d) QCBM, e) NLCD and f) Copula. The QCBM was trained on an ideal quantum simulator with inference performed on IonQ's Forte-Enterprise QPU. g) Threshold dependence of InSAR lava-flow change-detection performance. F1 score after uniform filtering as a function of the detection threshold for QCBM, NLCD and Copula on the InSAR lava-flow dataset. The ideal-simulator QCBM curve shows the mean over 10 independent training runs, with the shaded band indicating one standard deviation.}
    \label{fig:lava_flow_masks1}
\end{figure}

We next evaluated the methods on an InSAR dataset representing surface changes associated with a volcanic lava flow, shown in Figure~\ref{fig:lava_flow_masks1}(a)--(f). The selected QCBM used 11 bits per variable, corresponding to 22 qubits, and was trained using an ideal quantum simulator with inference performed on IonQ's Forte-Enterprise QPU. The best-performing Copula model used 12 bits per variable. QCBM, NLCD and Copula all localized the primary lava-flow region in the ground truth and achieved a maximum filtered F1 score of approximately 0.66. The QCBM mask obtained from QPU inference preserved the dominant changed region while exhibiting limited background response, and the three methods produced broadly similar masks in the principal lava-flow area.

The threshold dependence of the filtered F1 score is shown in Figure~\ref{fig:lava_flow_masks1}(g). The ideal-simulator QCBM curve represents the mean over 10 independent training runs, with the shaded band indicating one standard deviation. The ideal QCBM, QPU-inference QCBM, NLCD and Copula all reached a maximum filtered F1 score of approximately 0.66, indicating comparable peak performance on this InSAR dataset. The QCBM curves maintained high filtered F1 scores across a broad range of detection thresholds, providing a wider high-performance operating region than the classical counterparts over the relevant threshold range shown in Figure~\ref{fig:lava_flow_masks1}(g). The QPU-inference curve was shifted relative to the ideal-simulator QCBM curve and began to decrease at lower thresholds, indicating that execution on quantum hardware affected threshold calibration and operating range more than the maximum attainable filtered F1 score. This broad high-performance threshold region is practically relevant because the optimal threshold may not be known in advance and may vary across scenes or acquisition conditions.

Together, the SAR and InSAR results show that QCBM achieves performance comparable to or better than the classical baselines under the reported experimental conditions. The strongest relative gains occur for non-Gaussian SAR data, where the classical methods are less effective unless preprocessing regularizes the pixel-value distributions. On the InSAR lava-flow dataset, QCBM and the classical methods attain similar peak filtered F1 scores, while the QCBM threshold response indicates robustness over a broad operating range. These results support QCBM as a competitive generative model for image change detection across both simulator-based and QPU-inference settings.

\paragraph*{Quantum Sampling versus Classical Smoothing}

To assess whether the choice of conditional expectation estimator influences the comparison between the quantum and classical change-detection approaches, we performed an ablation study on the classical models NLCD and Copula. In their original implementations, NLCD and Copula estimate the background from the before $x$ and after $y$ images using a lookup-table estimator of the conditional expectation, $E[y|x]$.  As reported in Table \ref{table:1}, this approach yields filtered F1 scores of 0.16 and 0.24 for NLCD and Copula, respectively, on Miramar1-Med3 dataset. For the ablation study, the lookup-table estimator was replaced with kernel density estimation (KDE), while all other components of the change-detection pipelines were kept unchanged. Specifically, the \texttt{gaussian\_kde} implementation from
\texttt{scipy.stats} \cite{Scipy} was used to estimate the joint probability
density $p(x,y)$, from which the conditional expectation $E[y|x]$ was computed to generate the background. Under this KDE-based formulation, the best filtered F1 score obtained on dataset Miramar1-Med3 was only 0.05 for both NLCD and Copula, representing a substantial degradation relative to the corresponding lookup-table results. These findings demonstrate that, for dataset Miramar1-Med3, the KDE approach with a smoother, continuous estimate of the joint distribution does not improve the classical baselines and instead markedly reduces their change-detection performance. The ablation therefore supports the lookup-table formulation as the more effective conditional expectation estimator for NLCD and Copula on this dataset and indicates that their lower performance relative to QCBM cannot be attributed to the use of an inadequately simple conditional expectation estimator.

\paragraph*{Transfer across scenes without retraining.}

A limitation of the current quantum image change detection framework is that the model is typically trained independently for each before–after image pair, resulting in model parameters that are specific to the data used for training. Such per-image-pair optimization requires repeated training prior to inference and may therefore limit the practicality of the approach in deployment. To investigate whether the trained quantum model can generalize to previously unseen image data, we performed a cross-chip experiment using two spatially distinct chips from the same geographic region (Miramar airport in San Diego). The Original dataset reported in Table \ref{table:1} corresponds to the first chip (Miramar1), whereas the quantum model was trained exclusively using a second chip (Miramar2) and subsequently evaluated on the first chip without retraining or fine-tuning on the inference data. For this experiment, the before- and after-image pixel intensities were discretized using $n_{\mathrm{bits}}=10$ bits per variable, resulting in two 10-qubit registers and a total circuit size of 20 qubits. The model was trained on Miramar2 dataset using an ideal quantum simulator and then directly applied to Miramar1 (Original) for inference. Under this cross-chip setting, the quantum model achieved a final filtered F1 score of 0.27 on Miramar1. In comparison, the classical models NLCD and Copula achieved filtered F1 scores of 0.14 and 0.20, respectively, on the dataset, as reported in Table \ref{table:1}. Thus, despite being evaluated on data that were not used during training, the quantum model retained superior change-detection performance relative to both classical baselines. These results provide preliminary evidence that the learned quantum model parameters can transfer across image chips from the same geographic region, suggesting that per-image-pair retraining may not be strictly necessary and motivating further investigation of the model's generalization across spatial regions, acquisition conditions, and datasets.

\paragraph*{Sparsity of the conditional histogram as a resource.} 

Our results indicate that the QCBM, through the joint probability distribution it learns, performs comparably to the classical lookup-table methods when the marginal pixel-intensity distributions of the two images are approximately Gaussian, but substantially outperforms these methods when the pixel intensities are nonuniform and sparsely distributed across the available values. This performance gain persists when the lookup-table estimator is replaced by the KDE formulation considered here, indicating that the QCBM advantage is not eliminated by straightforward classical smoothing of the joint distribution. Furthermore, the quantum model continues to show a performance advantage when trained on a different image pair, providing evidence of its ability to generalize to unseen data. Quantum generative models have shown promise in learning from sparse data \cite{hibat2024framework} and have demonstrated the potential for improved generalization relative to classical machine-learning models \cite{gili2024generalization,gili2023quantum}. In our setting, the low occupancy of the pixel-intensity space represents a sparse training resource that may be difficult for classical methods to model and thus offers a potential avenue for quantum utility. Under this interpretation, we would expect the QCBM to exhibit its largest performance gains on the non-Gaussian image sets, for which the classical methods have insufficient statistics. For the Miramar1 Original, Original-Med3, NG-Med5, and NG-Med7 configurations, only \(1.0\%\), \(0.80\%\), \(4.6\%\), and \(4.14\%\) of the bins in the joint histogram of real-space pixel intensities are occupied, respectively. By contrast, \(17.1\%\) and \(12.7\%\) of the bins are occupied for the YJ-Med3 and YJ-Med7 configurations, respectively. Because the QCBM outperforms the classical models precisely in the cases where the joint distribution is most sparsely populated, these results support our conjecture that sparsity in the training data contributes to the QCBM’s relative performance advantage.

\section*{Discussion}

The QCBM matched or outperformed the classical estimators on both datasets. 
The degree of the improvement varied considerably between experiments, and the pattern of that variation provides an indication of where the method shows benefit.
In particular, the use of the QCBM was of higher importance on the preprocessing configurations with strongly skewed pixel distributions.
The beneficial behavior was not obvious after the Yeo-Johnson transformation, which made those distributions approximately
Gaussian, and it was also absent on the InSAR image change detection analysis, where all three methods reached the same peak filtered F1.
Another important observation, that fits the same picture was that the performance improved as the discretization became finer, leaving fewer observations in each cell of the conditional histogram.
The common factor across these cases is the occupancy of the conditional histogram, not a specific property of the radar imagery. 
The lookup-table estimator returns a conditional expectation only for cells that contain observations, and that estimate is noisy for cells containing only few. 
The trained model returns a conditional expectation for every cell, including those the observed data leave empty. 

The same pattern was observed when the model was run on the IonQ QPUs. 
Because training and inference can be moved to the QPU independently, their QPU costs can be assessed separately. 
In the case that we performed inference only in the QPU the F1 score recorded was slightly lower (by about $0.03$), regardless of how the model had been trained. 
Moving training to the QPU reduced the model performance by another $0.08$, regardless of where the inference was executed. 
Attributing these reduction of the model performance to a single factor is almost impossible. 
In the case of ideal training we used 100,000 measurement shots per iteration, while in the QPU training to reduce the training time we limited it to 5,000, so part of the loss can may come from sampling noise in the KL-divergence objective rather than from the device performance itself. 
Nonetheless, in all combinations of ideal- and QPU-based training/inference, including fully hardware-executed training and inference, remained well above NLCD at 0.16 and Copula at 0.24. 

On the InSAR scene all three methods reached the same maximum filtered F1, but the QCBM retained near-maximum F1 over a wider interval of detection thresholds. 
The threshold must be fixed before deployment and its optimum differs between scenes.
These results indicate the that QCBM was therefore significantly less sensitive both for the choice of threshold and to the choice of training data than the classical methods were. 

The model improvement can be explained from the performed ablation. 
Replacing the lookup table with kernel density estimation did not manage to recover the gap with the respective classical methods. 
This rules out the direct classical explanation, namely that the classical baselines were limited by on overly simplified conditional estimator. 

The advantage also grows with the size of the model. 
Filtered F1 rose from 0.09 at 4 qubits to 0.36 at 22 qubits, and over the same range the margin over the classical copula widened from approximately 0.06 to 0.12. 
The QCBM led the copula at every circuit size tested and overtook the lookup-table baseline at 12 qubits. 
Across the measured range, the trend motivates evaluation at larger circuit sizes, which both finer discretization and additional reference acquisitions would require.

Several aspects of the approach remain to be characterised. 
First, the transfer experiment established that a trained model can be applied an unseen chip without retraining. 
The natural next question arising is how far this extends across regions and acquisition conditions, and a larger collection of images would answer this question. 
Second, F1 values are reported as maxima over the detection threshold for all three methods, so the comparison is made on equal terms; the wider interval over which the QCBM sustains near-maximum F1 further suggests it would lose least when the threshold is set in advance. 
Third, the ablation ruled out kernel density estimation as an explanation for the gap. 
A broader survey of classical density estimators, including generative models with structural constrains (comparable to QCBM), would delimit the comparison even further and provide further insights on the better performance of the QCBM. 
Lastly, conditional sampling is currently performed by rejection, whose cost grows as the conditioning becomes even rarer, so more efficient conditioning schemes can possibly offer a direct route to extending our method further in the sparse regime, in which it is most effective. 

As on outlook of our work two extensions can follow. Conditioning on several reference acquisitions instead of one would raise the dimension of the joint distribution, emptying the histogram far faster than finer discretization does, while costing only linearly more qubits. 
For interferometric data, the quantity that carries the signal is a phase wrapped onto a circle, and the copula construction used here is not defined for such variables.
A circular formulation would allow phase and coherence to be modeled together rather than through intensity alone.

\section*{Methods}


\subsection*{Data and acquisition parameters} 

Two bi-temporal datasets were used in this study: a SAR amplitude pair and an InSAR pair.

\paragraph*{SAR amplitude dataset (Miramar1).}
The Miramar1 dataset consists of a bi-temporal pair of X-band acquisitions collected by Capella Space's Capella-2 (``Sequoia'') satellite, the first operational satellite of the Capella constellation \cite{castelletti2021capella}, over Marine Corps Air Station Miramar, San Diego, CA, USA, on 22 and 29 October 2020, at incidence angles of $32.89^{\circ}$ and $36.60^{\circ}$, respectively.
Both acquisitions were collected in Stripmap imaging mode with HH polarization at a nominal spatial resolution of 1.2\,m. A crop of $500m \times 500m$ covering the airfield and adjacent
built-up area was extracted from each geocoded image for analysis. Prior to analysis, the bi-temporal stack was despeckled with a refined Lee filter \cite{lee2017review} using a $3\times3$ window and coregistered with the GeFolki optical-flow algorithm \cite{plyer2015new}, which is well suited to high-resolution SAR scenes containing dense structural detail. This despeckling and coregistration pipeline follows that of our earlier unsupervised change-detection framework for very-high-resolution SAR \cite{de2021unsupervised}. 

\paragraph*{InSAR dataset (lava flow).}

The InSAR dataset represents surface change associated with the 2025--2026 eruptive sequence of Piton de la Fournaise (La R\'eunion Island, France), during which a major eruption beginning on 13 February 2026 produced lava flows that reached the Indian Ocean on 16 March 2026. The data were collected by Capella Space's Capella-13 satellite, which flies a $53^{\circ}$-inclined mid-inclination orbit with a 2.95-day repeating ground track, enabling interferometric pairs at a 3-day cadence \cite{staniewicz2026operational}. Acquisitions were collected in Stripmap imaging mode with HH polarization at X-band with a nominal spatial resolution of approximately $1\,\mathrm{m}$, over the summit and eastern flank, as part of a monitoring campaign conducted with
the Institut de Physique du Globe de Paris (IPGP) and the Observatoire Volcanologique du Piton de la Fournaise (OVPF) \cite{hauck2026capella}. 

The before and after images used for change detection are interferometric coherence magnitude maps. They were formed from two acquisitions, on 3 February, 6 February and 8 March 2026. The 3--6 February pair, which post-dates the 18 January north-flank eruption but pre-dates the 13 February eruption, forms the pre-event coherence image, while the 6 February--8 March pair spans the eruption and contains the lava flow. The stack was coregistered to sub-pixel accuracy and coherence was estimated from the resulting interferograms over a $4\times6$-pixel (range $\times$ azimuth) moving window, giving an approximately square multilooked ground cell. A crop covering the summit and the eastern flow field was extracted for analysis. Emplacement of fresh lava disrupts the scattering surface between passes, dropping coherence from near $1$ to near $0$, so the change signal in this dataset is a coherence loss rather than an amplitude change.

\paragraph*{Ground-truth masks.}
For the Miramar1 dataset, the ground-truth change mask was produced by manual expert annotation. For the lava-flow dataset, the ground-truth mask was manually annotated using external updates provided by OVPF/IPGP. All annotations were performed independently of the outputs of the algorithms evaluated in this work.


\subsection*{Preprocessing configurations}  

To evaluate the sensitivity of the quantum and classical estimators to the statistical character of the input data, six preprocessing configurations of the Miramar1 dataset were generated, differing in (i) the pointwise transformation applied to the pixel intensities and (ii) the size of a subsequently applied median filter. In each configuration, the same processing was applied identically to the before and after images. A configuration is denoted by its transformation followed by its median kernel; for example, YJ-Med3 denotes the Yeo--Johnson transformation followed by $3\times3$ median filtering.

\paragraph*{Original.} The despeckled, coregistered amplitude data with no further transformation. The single-look amplitude statistics of sub-meter X-band imagery leave these distributions strongly skewed and long-tailed (Figure~\ref{fig:miramar1_all_images}). 

\paragraph*{Median filtering (Med $k$).} A $k\times k$ median filter, which replaces each pixel with the median of its $k\times k$ neighborhood, was applied with $k\in\{3,5,7\}$. Median filtering suppresses isolated speckle-driven outliers while preserving edges, and largely preserves the
shape of the global intensity distribution; the Original-Med3 configuration therefore remains strongly non-Gaussian.

\paragraph*{Yeo--Johnson transformation (YJ).} The Yeo--Johnson power transform \cite{weisberg2001yeo} is a monotonic mapping defined for $\lambda\in\mathbb{R}$ as

\begin{equation*}
    \small
    y_{i}^{(\lambda)} =
     \begin{cases}
       \left[(y_i + 1)^{\lambda} - 1\right]/\lambda
         &\quad\text{if } \lambda \neq 0,\; y_i \geq 0,\\[2pt]
       \log(y_i + 1)
         &\quad\text{if } \lambda = 0,\; y_i \geq 0,\\[2pt]
       -\left[(-y_i + 1)^{2-\lambda} - 1\right]/(2-\lambda)
         &\quad\text{if } \lambda \neq 2,\; y_i < 0,\\[2pt]
       -\log(-y_i + 1)
         &\quad\text{if } \lambda = 2,\; y_i < 0,
     \end{cases}
\end{equation*}

with $\lambda = 1$ recovering the identity. The transform was applied using the \texttt{PowerTransformer} implementation in \texttt{scikit-learn} \cite{scikit-learn} with \texttt{method='yeo-johnson'}. A single value of $\lambda$ was estimated by maximum likelihood on the before and after images jointly, rather than fitting each image independently, so that both acquisitions are mapped through an identical pointwise transformation and their radiometric relationship, on which the joint distribution modelled by the estimators depends, is preserved. The optional post-transform standardization to zero mean and unit variance was disabled (\texttt{standardize=False}), leaving the transformed intensities on their native scale.

The transform substantially Gaussianizes the long-tailed amplitude distributions, and the YJ-Med3 and YJ-Med7 configurations (Yeo--Johnson followed by $3\times3$ and $7\times7$ median filtering) accordingly exhibit approximately Gaussian marginals (Figure~\ref{fig:miramar1_all_images}). These configurations represent the regime most favorable to the classical estimators.

\paragraph*{Non-Gaussian transformation (NG).} As a counterpart to the Yeo--Johnson configurations, the NG transformation is a fixed pointwise map
applied to accentuate rather than suppress the heavy tail of the amplitude distribution,

\begin{equation*}
    y_i^{\mathrm{NG}} = y_i^{\,p}, \qquad p = 1.2 \text{ (here)} > 1,
\end{equation*}

applied identically to both images with no per-image fitting, so that the resulting marginals remain strongly skewed and long-tailed by construction. Unlike the Yeo--Johnson transform, whose parameter is chosen to maximize the Gaussianity of the output, the NG map has no free parameter fitted to the data and therefore provides a controlled, reproducible stress case rather than a data-adaptive one. The NG-Med5 and NG-Med7 configurations apply this transformation followed by $5\times5$ and $7\times7$ median filtering, respectively, and so probe whether the estimators' behavior on heavy-tailed marginals persists at larger smoothing scales, where speckle has been substantially suppressed but the intensity distribution has not been regularized.

Table~\ref{table:preproc} reports all combinations of pre-processing applied to the datasets for amplitude change detection.

\begin{table}[h!]
\caption{Preprocessing configurations and kernel configuration referenced in
Table~\ref{table:1}. \\}
\centering
\small
\begin{tabular}{ l c c c c }
 \bf{Configuration} & \bf{Transform} & \bf{Kernel} \\
 \hline
 Original       & ---   & ---          \\
 Original-Med3  & ---   & $3\times3$    \\
 YJ-Med3        & YJ    & $3\times3$   \\
 YJ-Med7        & YJ    & $7\times7$   \\
 NG-Med5        & NG    & $5\times5$    \\
 NG-Med7        & NG    & $7\times7$   \\
 \hline
\end{tabular}
\label{table:preproc}
\end{table}

\subsection*{Nonlinear background estimation}  

Given a set of $M$ reference images, ${x_1,x_2,…,x_M}$, and an observed (after-event) image $y$, the Non-Linear Background Estimator (NLBE) aims to estimate the corresponding background image $\tilde{y}$ for change detection. For each pixel location $(i,j)$, the estimated background intensity is obtained by applying a nonlinear mapping $f$ to the pixel values from the reference images:

\begin{equation}
\tilde{y}(i,j) = f[x_1(i,j),x_2(i,j),...,x_M(i,j)],
\end{equation}
The function $f$ is chosen to minimize the mean squared estimation error $E[(\tilde{y}-y)^2]$ \cite{NLCD0}.

It has been shown in \cite{NLCD0} that the optimal estimator, in the minimum mean squared error (MMSE) sense, is the conditional expectation of $y$ given the reference images:
\begin{equation}
f = E[y|x_1,...,x_M],
\end{equation}

which can be expressed as
\begin{equation}\label{eq:3}
E[y|x_1,...,x_M] = \frac{\sum_y yp(x_1,...,x_M,y)}{\sum_yp(x_1,...,x_M,y)},
\end{equation}

where $p(x_1,...,x_M,y)$ is the joint probability distribution of the pixel values in the before images $x_1,...,x_M$ and the corresponding after image $y$.

In practice, these conditional expectation values can be efficiently computed using a lookup table. Once the estimated background image $\tilde{y}$ is obtained, a difference image is generated by calculating $|y-\tilde{y}|$. Applying an appropriate threshold to this difference image produces a binary change mask, which highlights regions where significant scene changes have occurred.

\subsection*{Copula representation} 

\subsubsection*{Introduction to Copulas}
Copulas provide a powerful framework for modeling the dependence structure among multiple random variables independently of their marginal distributions. This separation allows complex dependencies, including nonlinear and tail dependence, to be modeled while preserving the individual marginal distributions.

Given $n$ random variables $X_1,...,X_n$ with marginal cumulative distribution functions (CDFs) $F_1,...,F_n$, the corresponding copula variables are obtained through the probability integral transform,
\begin{equation}
(U_1,...,U_n) = (F_1(X_1),...,F_n(X_n)). \label{eq:copula_transform}
\end{equation}
This transform has the property that each new random variable, $U_i$, is uniformly distributed on the interval $[0,1]$. The copula $C$ describes the joint distribution of these transformed variables,
\begin{equation}
C(u_1,...,u_n) = P(U_1 \leq u_1,...,U_n \leq u_n). \label{eq:copula_distribution}
\end{equation}
Conversely, samples from the original variables can be recovered using the inverse marginal distributions, 
\begin{equation}\label{eq:copula_inverse_transform}
(x_1,...,x_n) = (F_1^{-1}(u_1),...,F_n^{-1}(u_n)).
\end{equation}

By Sklar's theorem \cite{Sklar}, the joint distribution of $X_1,...,X_n$ can be expressed as 
\begin{equation}
F(x_1,...,x_n) = C(F_1(x_1),...,F_n(x_n)),
\end{equation}
which separates the marginal distributions from their dependence structure. This property has made copulas widely used in finance, hydrology, machine learning, remote sensing, and image processing for modeling complex multivariate relationships.
\subsubsection*{Copula-Based Image Change Detection Algorithm}

This section introduces a copula-based image change detection algorithm that performs background estimation in the copula domain using the Non-Linear Background Estimator (NLBE). The algorithm takes as input $M$ reference images, an after image, and the number of discretization bits, $n_{bits}$.

The pixel intensities of the reference and after images are first linearly scaled to the interval $[-1,1]$. The scaled pixels are then transformed into the copula domain using the probability integral transform based on their marginal cumulative distribution functions (CDFs). The resulting copula samples are then discretized into integer values in the range $[0,2^{n_{bits}}-1)$.

The discretized reference images and the discretized after image are subsequently processed using the NLBE algorithm to estimate the background image in the discretized integer space. The estimated background is then mapped back to the continuous copula domain and transformed to the scaled intensity domain $[-1,1]$ using the inverse probability integral transform (see Eq.~\ref{eq:copula_inverse_transform}). Finally, the recovered image is rescaled to the original pixel intensity range to produce the estimated background image.

By performing background estimation in the copula domain, the proposed approach exploits the dependence structure of the image data while reducing the influence of the marginal intensity distributions, leading to a more robust background estimate for subsequent image change detection.

\subsection*{QCBM architecture}

Figure \ref{fig:miramar1_all_images} illustrates an example of a Quantum Circuit Born Machine (QCBM) circuit for modeling the joint distribution of $n=2$ random variables. The QCBM consists of two quantum registers, each containing $n_{bits}$ qubits used to encode one random variable. In general, modeling $n$ random variables requires a total of $n*n_{bits}b$ qubits. Each register is initialized by applying Hadarmard gates followed by CNOT gates that generate maximally entangled states. Parameterized unitary transformations, $U_i(\theta)$, are subsequently applied to each register. As shown in Figure \ref{fig:miramar1_all_images}, each unitary is composed of parameterized $R_z$, $R_x$, and $R_{ZZ}$ rotation gates, whose rotation angles collectively define the trainable parameter set $\theta$.

\subsubsection*{Training}
For image change detection, the QCBM copula is trained to learn the joint distribution of corresponding pixel intensities in reference and after images. The pixel values are first normalized to the interval $[-1,1]$ and transformed into copula space using their marginal cumulative distribution functions (CDFs). The resulting copula samples are discretized into binary representations using $n_{bits}$ bits per variable, producing a target bit string distribution, $P(x)$. Measurements of the QCBM generate samples from a model distribution, $Q(x)$. The circuit parameters are optimized by minimizing the Kullback–Leibler (KL) divergence between $P(x)$ and $Q(x)$ using a classical optimization algorithm such as Constrained Optimization BY Linear Approximation (COBYLA) \cite{Powell1994}.

\subsubsection*{Conditional sampling} 

After the QCBM circuit is trained, samples are generated by repeatedly measuring the quantum circuit. Each measurement produces a bit string of length $nn_{bits}$, which is partitioned into $n$ sub strings of length $n_{bits}$, with each sub string representing one random variable. These sub strings are decoded into samples in copula space, denoted by $u$. The copula samples are then transformed into samples from the original joint distribution, $F(x_1,…,x_n)$, by applying the inverse marginal cumulative distribution functions to each variable.

The background image is estimated by drawing samples from the learned joint distribution conditioned on the observed pixel intensities of the reference image. This conditional sampling is performed using rejection sampling, whereby candidate samples are repeatedly drawn until they satisfy the specified conditioning constraints. Samples are generated from the trained QCBM until conditional distributions corresponding to all relevant reference-image pixel intensity combinations are obtained. The estimated background pixel intensity is then computed as the expectation (or sample mean) of each conditional distribution, analogous to the nonlinear background estimation (NLBE) method. Unlike NLBE, however, the conditional distributions are obtained from samples generated by the learned quantum copula (QCBM) rather than directly from the observed data.

\subsection*{Quantum hardware execution}
To evaluate the proposed quantum image change detection model (QCBM) under realistic hardware conditions, both training and inference were performed on a quantum processing unit (QPU) using Miramar1-Med3 dataset. The before and after image intensities from the dataset were discretized using $n_{\mathrm{bits}}=10$ bits, and the corresponding quantum circuit employed two $n_{\mathrm{bits}}$-qubit registers, one for each image, resulting in a total of $2n_{\mathrm{bits}}=20$ qubits. The QPU-executed circuit contained 50 single-qubit gates and 28 two-qubit gates. Whereas the ideal-simulation experiments employed the COBYLA optimizer \cite{Powell1994}, QPU training used simultaneous perturbation stochastic approximation (SPSA) \cite{SPSA}, which is well suited to optimization in the presence of finite-shot sampling and hardware noise. The SPSA learning-rate and perturbation parameters were both initialized to 0.3. Training on Miramar1-Med3 dataset was performed for 300 optimization iterations, with 5,000 measurement shots used for each circuit evaluation. In comparison, the ideal-simulation experiments were trained for 1,000 iterations; the QPU training budget was reduced to 300 iterations to limit the number of quantum circuit evaluations and associated hardware execution cost. Following training, inference on the dataset was also performed directly on the QPU, using 100,000 measurement shots to obtain a higher-statistics estimate of the output probability distribution used for image change detection. This experiment therefore evaluates the complete training and inference pipeline on quantum hardware using a 20-qubit implementation, rather than restricting hardware execution to inference from parameters optimized under ideal simulation.

\subsection*{Classical baselines and ablations}
For the ablation study, the lookup-table estimator used by the classical methods was replaced by a kernel density estimate of the joint distribution of the before-image and after-image pixel intensities. Let \(\{(x_i,y_i)\}_{i=1}^{N}\) denote the paired pixel intensities from the before image \(X\) and after image \(Y\), respectively. The joint density \(p(x,y)\) was estimated using the \texttt{gaussian\_kde} implementation in \texttt{scipy.stats}\cite{Scipy}, with the paired observations supplied as a \(2\times N\) dataset. The resulting KDE can be expressed as an equally weighted mixture of bi-variate Gaussian kernels,
\begin{equation}
p(x,y)=\frac{1}{N}\sum_{i=1}^{N}
\mathcal{N}\!\left(
\begin{bmatrix}x\\y\end{bmatrix};
\begin{bmatrix}x_i\\y_i\end{bmatrix},
\boldsymbol{\Sigma}
\right),
\end{equation}
where \((x_i,y_i)\) is the center of the \(i\)th kernel and \(\boldsymbol{\Sigma}\) is the common kernel covariance matrix determined by \texttt{gaussian\_kde}. The covariance matrix is available from the fitted KDE through its \texttt{covariance} attribute and can be written as
\begin{equation}
\boldsymbol{\Sigma}=
\begin{bmatrix}
\sigma_x^2 & \sigma_{xy}\\
\sigma_{xy} & \sigma_y^2
\end{bmatrix},
\end{equation}
where \(\sigma_x^2\) and \(\sigma_y^2\) are the kernel variances in the before- and after-image dimensions and \(\sigma_{xy}\) is their covariance. For a given before-image intensity \(X=x\), the conditional expectation required by the classical background model is
\begin{equation}
E[Y|X=x]
=\int y\,p(y|x)\,dy.
\end{equation}
Because \texttt{gaussian\_kde} represents the joint density as a Gaussian mixture, this integral can be evaluated analytically. The relative contribution of the \(i\)th kernel at \(X=x\) is
\begin{equation}
w_i(x)=
\exp\!\left[-\frac{(x-x_i)^2}{2\sigma_x^2}\right],
\end{equation}
and the conditional expectation implied by the fitted KDE is therefore
\begin{equation}
E[Y|X=x]
=
\frac{
\displaystyle\sum_{i=1}^{N}
w_i(x)
\left[
y_i+\frac{\sigma_{xy}}{\sigma_x^2}(x-x_i)
\right]
}{
\displaystyle\sum_{i=1}^{N}w_i(x)
}.
\end{equation}
Thus, $E[Y|X=x]$ is a weighted combination of the conditional means of the individual Gaussian kernels, with larger weights assigned to samples whose before-image intensities $x_i$ are closer to the query intensity $x$. All quantities required for this calculation are obtained directly from the fitted \texttt{gaussian\_kde} model: the paired samples $(x_i,y_i)$ are contained in its \texttt{dataset} attribute, while $\sigma_x^2$ and $\sigma_{xy}$ are obtained from the corresponding elements of its \texttt{covariance} matrix. The KDE-derived conditional expectations were then used in place of the lookup-table estimates of $E[Y|X]$ to construct the background images for the classical methods, allowing the effect of the conditional expectation estimator to be isolated in the ablation study.

\subsection*{Evaluation} 

The performance of the quantum and classical image change detection models was evaluated using the F1 score of the resulting binary change masks. For each model, the predicted background image, $\hat{B}$, was first compared with the after image, $Y$, to construct a pixel-wise absolute difference image, $D=|\hat{B}-Y|$. Because the input images are 16-bit, candidate binary change masks were generated by thresholding $D$ over pixel-intensity thresholds in the range $[0,65535]$, with pixels exceeding a given threshold classified as changed and the remaining pixels classified as unchanged. The predicted masks were compared with the ground-truth change masks using the F1 score, defined as the harmonic mean of precision and recall,

\begin{equation}
F_1 = 2\frac{\mathrm{Precision}\times\mathrm{Recall}}
{\mathrm{Precision}+\mathrm{Recall}}
=\frac{2TP}{2TP+FP+FN},
\end{equation}

where $TP$, $FP$, and $FN$ denote the numbers of true-positive, false-positive, and false-negative pixels, respectively. To reduce isolated detections and improve the spatial consistency of the predicted masks, each binary mask was subsequently processed using the \texttt{uniform\_filter} function from \texttt{scipy.ndimage}\cite{Scipy}. Uniform-filter window sizes from 3 to 11 pixels were considered, and the filtered response was thresholded using percentage thresholds from $10\%$ to $100\%$ in increments of $10\%$. Thus, for each pixel-intensity threshold, all combinations of uniform-filter size and percentage threshold were evaluated, and the maximum F1 score across these combinations was retained as the filtered F1 score for that pixel threshold. Finally, the overall filtered F1 score reported for each model was defined as the maximum filtered F1 score obtained across all evaluated pixel-intensity thresholds. This procedure jointly accounts for the intensity threshold used to distinguish change from background and the spatial consistency of the detected regions, enabling a consistent comparison of change-detection performance across the quantum and classical models.

\section*{Acknowledgment} 
The authors would like to acknowledge Justin Rodriguez for providing expertise in product pipelines and road maps that guided the scope of this project. The authors would also like to acknowledge Jonathan Mei for laying foundations for the classical copula method.

\section*{Author Contributions}
S.K.S and E.E. developed the theoretical quantum framework. S.D., S.S., and C.S. contributed expertise on SAR, InSAR, and classical methods. S.K.S. executed the experiments. S.D and S.S. supplied the SAR/InSAR datasets and generated the corresponding ground truths. P.K.B., A.H., M.R., and J.I. supervised the work. All authors analyzed results and contributed to the preparation of the manuscript.

\section*{Correspondence}
Correspondence and requests for material should be addressed to Samwel K. Sekwao

\clearpage

\bibliographystyle{IEEEbib}
\small
\bibliography{References/changedet}

\end{document}